\documentclass[11pt,a4paper]{article}
\usepackage{jheppub}

\title{Brane polarization is no cure for tachyons}

\author{Iosif Bena}  \author{and Stanislav Kuperstein}

 \affiliation{Institut de Physique Th\'eorique, Universit\'e Paris Saclay, CEA, CNRS, F-91191 Gif-sur-Yvette, France}

\emailAdd{iosif.bena@cea.fr} \emailAdd{stanislav.kuperstein@gmail.com}

\abstract{Anti-M2 and anti-D3 branes placed in regions with charges dissolved in fluxes have a tachyon in their near-horizon region, which causes these branes to repel each other. If the branes are on the Coulomb branch this tachyon gives rise to a runaway behavior, but when the branes are polarized into five-branes this tachyon only appears to lower the energy of the polarized branes, without affecting its stability. We analyze brane polarization in the presence of a brane-brane-repelling tachyon and show that when the branes are polarized along the direction of the tachyon the polarized shell is unstable. This implies that tachyons cannot be cured by brane polarization and indicates that, at least in a certain regime of parameters, anti-D3 branes polarized into NS5 branes at the bottom of the Klebanov-Strassler solution have an instability. }

\preprint{IPhT-T15/038}

\begin{document}
\maketitle

\setcounter{footnote}{0}
\setcounter{figure}{0}
\setcounter{equation}{0}

\section{Introduction}

The physics of antibranes placed in backgrounds that contain charges dissolved in the fluxes has been a constant source of surprises. The supergravity solutions that describe these antibranes have certain singularities, that are visible both when the fields sourced by the branes are treated as a perturbation \cite{Bena:2009xk, Bena:2011hz, Dymarsky:2011pm, Bena:2011wh,Bena:2010gs,Giecold:2013pza} and when the full backreaction of these fields is taken into account \cite{Blaback:2011nz,  Blaback:2011pn, Bena:2012bk,  Gautason:2013zw}. The fields diverging at the singularity have the right form and $g_s$ dependence to cause the antibranes to polarize into higher-dimensional branes, but in all regimes of parameters where the exact coefficients of these fields could be computed they are such that polarization either does not happen \cite{Bena:2012tx, Bena:2012vz}, or if it happens it is accompanied by  a brane-brane-repelling tachyon, which causes the antibranes to run away from each other \cite{Bena:2014bxa, Bena:2014jaa}\footnote{When the worldvolume of the branes is not flat but AdS, this singularity can be resolved by polarization \cite{Apruzzi:2014qva, Junghans:2014wda}, but this can only happen when the scale of the AdS space is the same as the scale of the polarizing fields \cite{Bena:2012tx}. Hence, this phenomenon is irrelevant for resolving the singularities of the antibranes that are added to AdS flux compactifications to obtain de Sitter spaces with small cosmological constant \cite{Kachru:2003aw}.}.

The generic potential for $n$ anti-D3 branes to polarize into $N_5$ five-branes wrapping a two-sphere of radius $R$ in a three-form field strength proportional to $C$ is of the form \cite{Polchinski:2000uf}:
\begin{equation}
\label{eq:TheStandardPotential}
V_{\rm polarization} = \frac{N_5^2}{n} R^4 - C N_5  R^3 - m^2 n R^2 \, . 
\end{equation}
The first term represents the excess mass brought about by the presence of the five-branes, the second term is the polarization force exerted by the six-form potential (Hodge dual to three-form field strength) on the five-branes, and the last term is the potential felt by the $n$ anti-D3 branes. When a tachyon is present ($m^2 >0$) the polarization potential has always a minimum, regardless of the sign a $C$. This implies that anti-D3 branes can polarize into five-branes even when the cubic term in the polarization potential  (and hence the six-form potential that induces the branes to polarize) is absent or has the wrong sign \cite{Zamora:2000ha}. Hence, the presence of a brane-brane-repelling tachyon appears to favor brane polarization. One may even go so far as to argue that the presence of this tachyon should be taken as an indication that antibrane singularities will always be resolved by brane polarization, and that the tachyon simply signals the intention of the unstable unpolarized anti-D3 branes to move to a stable vacuum where the antibranes are polarized into five-branes. 

It is the purpose of this paper to investigate the stability of the vacua with polarized antibranes in the presence of a brane-brane-repelling tachyon.
%and to show that these vacua are unstable
We will examine polarized branes with a tachyonic term in the polarization potential and examine two possible instability modes, both of which break the spherical symmetry of the polarization shell: the P-mode, which corresponds to shifting the center of the polarization shell, and the D-mode, which corresponds to deforming the spherical polarization shell into a ellipsoidal (cucumber) shape. We will find that the P-mode deformation is always tachyonic, while the D-mode is only tachyonic when the three-form field strength that polarizes the branes is smaller than a certain value ($C < \frac{2}{3} m$). Hence, the presence of a brane-brane-repelling tachyon renders unstable the configurations where these branes are polarized into higher-dimensional branes.

The paper organized as follows. In Section \ref{sec:Action} we write down the potential describing the polarization of D3 branes into a five-brane wrapping a (shifted and  squashed) non-spherical polarization shell. The next two sections are devoted to the stability analysis of this shell: In Section \ref{sec:P-wave} we study the stability under a P-wave deformation, corresponding to a shift of the center of the sphere, and find that this deformation is tachyonic whenever a tachyon is present in the polarization plane, regardless of the magnitude of the charges or of the polarizing fields. In Section \ref{sec:D-wave} we study the stability under a D-wave (ellipsoidal) deformation and show that this deformation is also tachyonic whenever the tachyon is stronger than a certain value. In the last section we discuss possible extensions and generalizations of our results and their implications for the stability of anti-D3 branes polarized into NS5 branes \cite{Kachru:2002gs} at the bottom of the Klebanov-Strassler solution \cite{Klebanov:2000hb}. Some lengthy formulae are relegated to the Appendix.

\section{The polarization potential of a non-spherical shell}
\label{sec:Action}

The goal of this section is to derive  the polarization potential for D3 branes polarized into five-branes that wrap a two-sphere perturbed with P- and D-mode deformations inside an $\mathbb{R}^3$ along which the inter-D3 potential is tachyonic.  
The five-brane worldvolume is spanned by the four Minkowski coordinates of the D3 branes and two compact directions inside the  $\mathbb{R}^3$ parameterized as:
\begin{eqnarray}
\label{eq:parametrization}
x_1 &=& R_1 \sin \theta \cos \varphi
\nonumber \\
x_2 &=& R_1 \sin \theta \sin \varphi
\\
x_3 &=& R_2 \cos \theta + a \,.
\end{eqnarray}
When the shift parameter $a$ is non-zero and the radii are equal this embedding describes a P-wave deformation of the spherical shell. The D-wave deformation corresponds to $a=0$ and $R_1\neq R_2$. The pull-back of the $\mathbb{R}^3$ metric on the deformed two-sphere wrapped by the five-brane is:
\begin{equation}
g_\perp = \left( R_1^2 \cdot \frac{\xi^2}{1-\xi^2} + R_2^2 \right) \textrm{d} \xi^2 + R_1^2 ( 1 - \xi^2 ) \textrm{d}  \varphi^2 \, ,
\end{equation}
where we introduced:
\begin{equation}
\label{eq:xi}
\xi \equiv \cos \theta \, .
\end{equation}
The D3 brane charge of the five-brane gives rise to a nontrivial 2-form flux on its worldvolume:
\begin{equation}
\label{eq:2-form}
F_2 = n f(\xi) \textrm{d} \xi \wedge \textrm{d} \varphi \, ,
\end{equation}
where $n$ is the total number of D3 branes and $ f(\xi)$ is the D3 charge density function normalized as:
\begin{equation}
\label{eq:constraint}
\int_{-1}^{1}  f(\xi) \textrm{d} \xi = 1 \, .
\end{equation}
For the spherically symmetric configuration $f(\xi) = \frac{1}{2}$, but when the sphere is deformed $f(\xi)$ will change, and has to be determined anew. To ensure that the total D3 brane charge is the same, one has to add a global ($\xi$-independent) Lagrange multiplier term to the probe action:
\begin{equation}
\label{eq:Lagrange-term}
\lambda \cdot \left(  f(\xi) - \frac{1}{2} \right) \, .
\end{equation}

As explained in \cite{Polchinski:2000uf}, the quartic term in the polarization potential is equal to the mass difference between $n$ unpolarized D3 branes and $n$ D3 branes inside a five-brane shell. When the total mass of the D3 branes is bigger than the five-brane mass this is proportional to:
\begin{equation}
\label{eq:quartic-term}
\frac{\textrm{det} g_\perp}{4 \sqrt{\textrm{det} F_{2}}} = \frac{1}{4 n f(\xi)}  \left( R_1^2  \xi^2 + R_2^2 \left( 1 -  \xi^2 \right)  \right) R_1^2 \, ,
\end{equation}
where the factor of 4 in the denominator was chosen such that the integral of this term matches the first term in (\ref{eq:TheStandardPotential}).

The cubic term in the polarization potential (\ref{eq:TheStandardPotential}) comes from the six-form potential that forces the five-brane shells to expand. In the near-brane region the field strength corresponding to this potential is proportional to the volume form of the $ \mathbb{R}^3$ plane in which the polarization happens: 
\begin{equation}
\label{eq:vol-form}
d C_6 \sim  \textrm{d} x_1 \wedge  \textrm{d} x_2 \wedge  \textrm{d} x_3   \wedge \textrm{Vol}_{\textrm{Mink}} \, ,
\end{equation}
%
%
%originates from the closed and co-closed 3-form \cite{Polchinski:2000uf} defined by $\omega_3^+ = h^{-1} \left( \star_6 G_3 + i G_3 \right)$, where $h$ is the D3 warp factor and $G_3$ is the anti-self dual\footnote{Notice that by definition $\omega_3^+$ does not depend on the self-dual part of $G_3$.} complex 3-form, the supergravity dual of the gauge theory fermionic masses. The $\textrm{d} \omega_3^+ = \textrm{d} \star_6 \omega_3^+ = 0$  condition it follows that $\omega_3^+ \sim  \textrm{d} x_1 \wedge  \textrm{d} x_2 \wedge  \textrm{d} x_3$.
%
Integrating the pullback of $C_6$ over the worldvolume of the five-brane gives the potential density:
\begin{equation}
\label{eq:cubic-term}
- \frac{3}{2} C R_1^2 R_2 \xi^2 \, ,
\end{equation}
where $C$ is a constant proportional to the strength of the polarizing fields, the factor of $3/2$ was chosen for later convenience and the minus sign reflects the fact that the orientation of the five-brane is such that the force exerted by the six-form potential favors polarization.

The quadratic term in the polarization potential is given by the D3 brane density multiplied by the value of the tachyonic potential at their location:
\begin{equation}
\label{eq:quadratic-term}
- m^2 n f(\xi) \left( x_1^2 +  x_2^2 + x_3^2 \right) \, ,
\end{equation}
where the minus sign reflects the presence of a brane-brane-repelling tachyon and the expression  $x_1^2 +  x_2^2 + x_3^2$ has to be pulled back on the worldvolume using the parameterization (\ref{eq:parametrization}). When the number of five-branes is larger than one, the full potential can be obtained from the potential above by simply replacing $n$ by $n /N_5$ and multiplying by an overall factor of $N_5$.
 
\section{The P-wave instability}
\label{sec:P-wave}

\begin{figure}[t]
\centering
\includegraphics[trim = 0mm 130mm 0mm 30mm, scale=0.4]{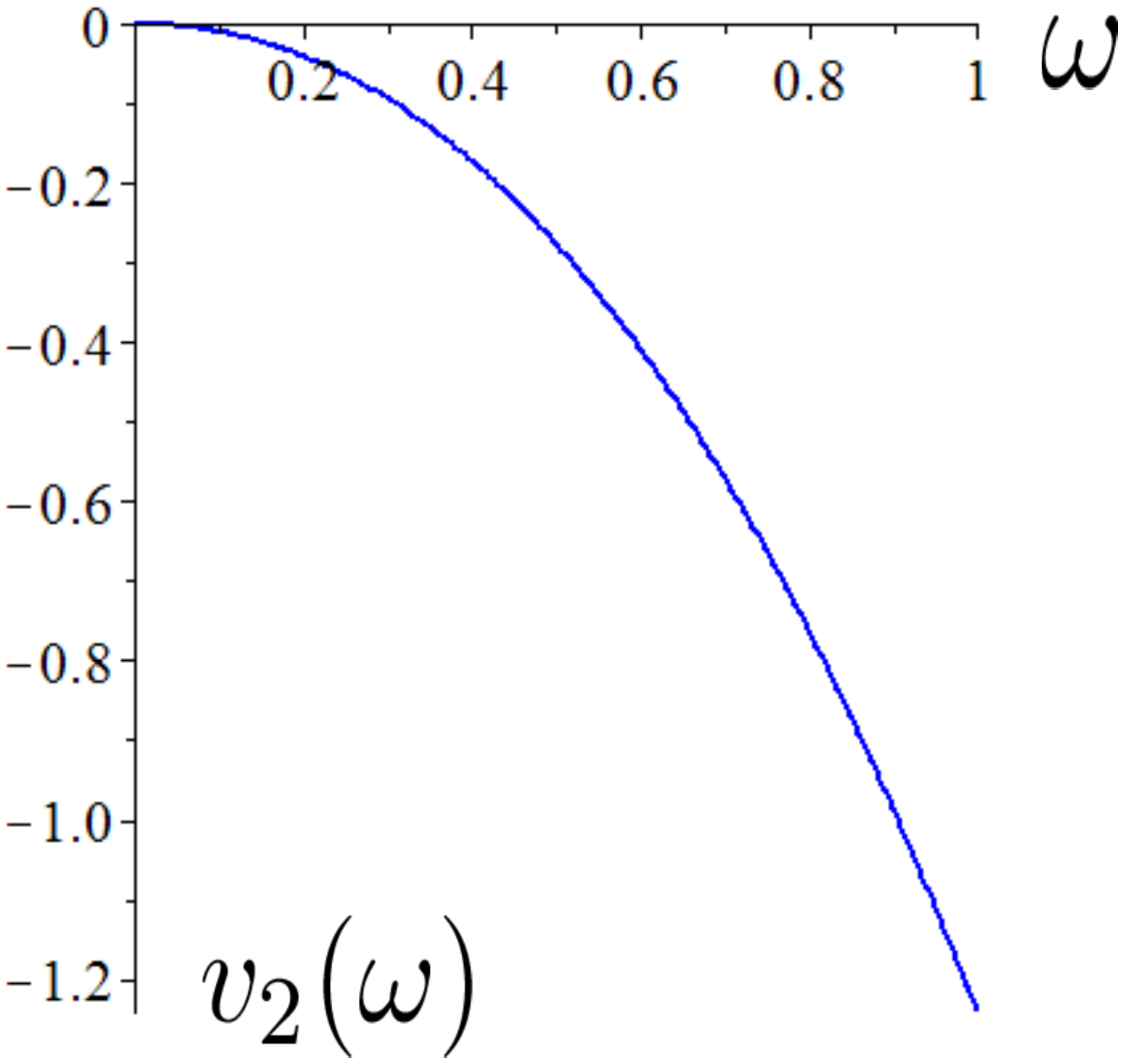}
\caption{The graph shows the $a^2$ term of the polarization potential $v_2(\omega)$ (with $\omega \equiv m/C$), which is always negative.}
\label{fig:V2}
\end{figure}

In this section we are interested in the polarization potential into a five-brane wrapping a shifted sphere given by (\ref{eq:parametrization}) with $R_{1,2}=R$ and non-zero $a$. With this parametrization we find that $x_1^2+x_2^2+x_3^2= R^2+2 R a \xi +a^2$ and, using equations (\ref{eq:Lagrange-term}), (\ref{eq:quartic-term}), (\ref{eq:cubic-term}) and (\ref{eq:quadratic-term}), we obtain the potential density:
\begin{equation}
\label{eq:Potentail-Shifted}
\mathcal{V} \left( f(\xi), \xi, R, a, \lambda \right) =  - m^2 n f(\xi) \left(  R^2+2 R a \xi +a^2 \right) -\frac{3}{2} C R^3 \xi^2 + \frac{ R^4  }{4 n f(\xi)} +  \lambda \cdot \left(  f(\xi) - \frac{1}{2} \right) \, .
\end{equation}
To find the final potential, $V(R,a) = \int_{-1}^1 \mathcal{V} \left( f(\xi), \xi, R, a, \lambda \right) \textrm{d} \xi $ one has to solve 
\begin{equation}
\label{eq:f-EOM}
\frac{\delta \mathcal{V} }{\delta f(\xi)} = 0 \, ,
\end{equation}
together with the constraint (\ref{eq:constraint}). Unfortunately, for a non-zero $a$ the solution of (\ref{eq:f-EOM}) for the charge distribution function,
\begin{equation}
\label{eq:f-Sol-Shifted}
f(\xi) = \dfrac{R^2}{2}  \left( \lambda n  - m^2 n^2 \left( R^2 + 2 R a \xi + a^2 \right) \right)^{-1/2} \, ,
\end{equation}
has a complicated dependence on $\xi$ which does not yield an analytic final expression for $\lambda$ and consequently for the potential $V(R,a)$.\footnote{Note that when $a=0$ the function $f(\xi)$ is $\xi$-independent, as expected for a spherically-symmetric configuration.}
Nevertheless, we are interested only in the small-$a$ behavior of the potential, and so it will suffice to solve for the Lagrange multiplier in a series expansion:
\begin{equation}
\lambda (a) = \lambda_0 + \lambda_1 a + \lambda_2 a^2 + \ldots \, ,  
\end{equation}
where the coefficients depend on $m, n$ and $R$. Expanding (\ref{eq:f-Sol-Shifted}) in $a$ and integrating term by term in $\xi$ we find that (\ref{eq:constraint}) implies:
\begin{equation}
\lambda_0 = \frac{R^4}{n} \left( 1 + \left( \frac{m n}{R} \right)^2 \right) \, ,
\qquad
\lambda_1 = 0 \, ,
\qquad
\lambda_2 = m^2 n \left( 1 + \left( \frac{m n}{R} \right)^2 \right) \, 
\end{equation}
and for the purpose of this analysis we will not need higher-order coefficients. Substituting this expansion into (\ref{eq:f-Sol-Shifted}) and (\ref{eq:Potentail-Shifted}) and integrating the potential density over $\xi$, we arrive at the final expression for $V(R,a)$:
\begin{equation}
\label{eq:V(R,a)}
V(R,a) = \left( - m^2 n R^2 -C R^3 + \frac{R^4}{n} \right) - m^2 n  \left( 1 + \frac{1}{3} \left( \frac{m n}{R} \right)^2 \right) \cdot a^2 + \mathcal{O}\left( a^3 \right) \, .
\end{equation}
The coefficient of the $a^2$ term is strictly negative, but this does not yet imply P-wave instability. Indeed, we have to expand the potential around the polarization radius $R=R_\textrm{pol}$ which minimizes the full potential including the  $a^2$ term:
%\footnote{Note that this there is also a solution with a negative value of $R_\textrm{pol}$. This, however, corresponds to the opposite orientation of the probe brane.}
\begin{eqnarray}
\label{eq:Rp}
R_\textrm{pol} &=& \dfrac{n}{8} \left( 3 C + \sqrt{9 C^2+32 m^2} \right) 
-\frac{1024}{3} \dfrac{m^4}{n} \Big[ C  \left( 243 C^4 + 1296 C^2 m^2 +  1536 m^4 \right) 
\nonumber \\
&& \qquad 
+ \left( 81 C^4 + 288 C^2 m^2 +128 m^4  \right) \sqrt{9 C^2+32 m^2} \Big]^{-1} \cdot a^2 + \ldots \, . 
\end{eqnarray}
Substituting $R=R_\textrm{pol}$ into (\ref{eq:V(R,a)}) one arrives at the final form of the potential we are interested in:
\begin{equation}
\label{eq:V-onshell}
V_\textrm{on-shell} (n,C,m) = n^3 C^4 v_0 (\omega) + n C^2 v_2  (\omega) \cdot a^2 + \mathcal{O} \left( a^3 \right)\, ,  \qquad \textrm{where} \qquad
\omega \equiv \frac{m}{C} \,\, , 
\end{equation}
and the full expressions for $v_0  (\omega)$ and $v_2 (\omega)$ are given in the Appendix. The leading term, $v_0(\omega)$, is the value of the potential for a spherically symmetric ($a=0$) configuration, and one can see from equation (\ref{eq:v_2app}) and from Figure \ref{fig:V2} that the term quadratic in $a$ is always negative, regardless of the values of $m$, $C$ and $n$.

Hence, a small shift of the center of the polarized sphere lowers its potential energy and makes this configuration P-wave unstable.

\section{The D-wave instability}
\label{sec:D-wave}

In this section we will study a D-wave (ellipsoidal) deformation of the spherical shell. When $a=0$, equation (\ref{eq:parametrization}) implies that $x_1^2+x_2^2+x_3^2= R_1^2 (1-\xi^2) + R_2^2 \xi^2$, and using (\ref{eq:Lagrange-term}), (\ref{eq:quartic-term}), (\ref{eq:cubic-term}) and (\ref{eq:quadratic-term}) we obtain:
\begin{eqnarray}
\label{eq:Potentail-Ellipsoid}
\mathcal{V} \left( f(\xi), \xi, R_1, R_2, \lambda \right) &=&  - m^2 n f(\xi) \left(  R_1^2 (1-\xi^2) + R_2^2 \xi^2 \right) -\frac{3}{2} C R_1^2 R_2 \xi^2 +
\nonumber \\
&&
+ \frac{R_1^2}{4 n f(\xi)} \left( R_1^2 \xi^2 + R_2^2 (1-\xi^2) \right)+  \lambda \cdot \left(  f(\xi) - \frac{1}{2} \right) \, .
\end{eqnarray}
As in the previous section, we will consider an infinitesimal deviation from a spherical shell:
\begin{equation}
R_1 = R + \sigma \cdot r_1 \, , \qquad R_2 = R + \sigma \cdot r_2 \, ,
\end{equation}
where $\sigma$ is a (dimensionless) ``squashing" parameter, and $R$ is the radius of the undeformed sphere (given by the leading-order ($a=0$) term in (\ref{eq:Rp})).

Proceeding as earlier, we solve (\ref{eq:f-EOM}) to find the D3 charge density distribution, $f(\xi)$, subject to the constraint (\ref{eq:constraint}). This allows us to obtain the perturbative $\sigma$-expansion for the Lagrange multiplier $\lambda$.  For our purposes it is enough to determine $\lambda$ only up to ${\cal O}(\sigma^2)$, which upon substituting back to the potential gives us the following quadratic term:
\begin{equation}
n C^2 \left( b_1 (\omega) r_1^2 + b_2 (\omega) r_2^2 + b_3 (\omega) r_1 r_2 \right) \cdot \sigma^2 \, ,
\label{eq:elipsoid-pot}
\end{equation}
where the expressions for $b_i(\omega)$ are given in the Appendix. The spherical solution is then stable with respect to D-wave fluctuations if and only if both eigenvalues of the corresponding matrix are positive. The product of the eigenvalues is  $4 b_1 b_2 - b_3^2$, and this expression is zero when
\begin{equation}
m = \dfrac{3}{2} C \, .
\end{equation}
It is straightforward to check that when the tachyon is larger\footnote{Recall that in our conventions $m$ and $n$ are positive.} than this critical value, $m > \frac{3}{2} C$, one of the eigenvalues is always negative, and hence the configuration is unstable. In particular, all the polarization channels where the polarizing fields are zero or negative ($C \leq 0$) and the polarization only happens because of the tachyon \cite{Zamora:2000ha}, give rise to configurations that are D-wave unstable. However, when the polarizing field is positive and much stronger than the tachyon, the D-wave instability is absent.

\section{Discussion}
\label{sec:Discussion}

We have found that brane polarization in the presence of brane-brane-repelling tachyons results in unstable configurations. Our analysis  focused of the stability of polarized branes that lay inside an $\mathbb{R}^3$ plane along which there is a brane-brane-repelling tachyon, and the perturbations that we explore do not take the polarization shell outside of this plane. However, in general a D3 brane solution perturbed with transverse three-form fields can have many polarized vacua, corresponding to polarization into arbitrary $(p,q)$ five-branes wrapping two-spheres laying in various $\mathbb{R}^3$ subspaces of the $\mathbb{R}^6$ transverse to the D3 branes \cite{Polchinski:2000uf}. Hence, in order to show that tachyons are incompatible with metastable vacua with polarized branes one must in principle analyze the stability of all such polarization shells,
% with arbitrary $(p,q)$ five-brane charges laying in all possible $\mathbb{R}^3$ subspaces, 
and consider more general perturbations than the ones we have considered in this paper. While we leave the thorough analysis of the stability of these configurations to a subsequent paper \cite{tilted-cucumber}, we outline below a few aspects of this analysis that reveal that the instability we found may be fatal for all brane polarization channels, regardless on whether they are along a plane where there is a brane-brane-repelling tachyon. 

In order to do this, it is important to recall that in the supersymmetric Polchinski-Strassler dual of the ${\mathcal N}=1^*$ gauge theory \cite{Polchinski:2000uf}, the polarization of the D3 branes into $(p,q)$ five-branes only gave rise to a supersymmetric vacuum when the polarization two-sphere was inside a specific $\mathbb{R}^3$, where the cubic term in the polarization potential is maximal (since the cubic term comes from the polarization force exerted on the shell, we can call this the ``maximum-force'' $\mathbb{R}^3$). However, nothing stops one from considering D3 branes polarized into $(p,q)$ five-branes wrapping a two-sphere inside a different $\mathbb{R}^3$ plane, such as a D5 shell in an oblique plane, or an oblique shell in the D5 or NS5 planes. These polarized shells will have more energy than the supersymmetric vacua, and are unstable under deformations that keep the polarized shell inside the same $\mathbb{R}^3$ plane. However, if we tilt these shells towards the ``maximum-force'' $\mathbb{R}^3$, their energy is lowered, and hence these shells are not even sitting on extrema of the polarization potential. 

When supersymmetry is broken there is no longer a natural way to associate a given type of $(p,q)$ five-brane to a given polarization $\mathbb{R}^3$ subspace of $\mathbb{R}^6$. Since both the cubic and the quadratic terms depend on the orientation of the polarization plane, one can also identify, besides the ``maximum-force'' $\mathbb{R}^3$, a ``maximum-tachyon'' $\mathbb{R}^3$, along which the quadratic term in the potential is the largest.\footnote{See for example equation (2.11) in \cite{Zamora:2000ha} or equation (3.1) in  \cite{Bena:2014jaa}.}
In general these two three-planes will not be the same, and the minimum-energy round shell will lay inside an $\mathbb{R}^3$ plane between them.  If there is a brane-brane-repelling tachyon in this plane our result implies that this round shell will be unstable. However, if this plane has a positive quadratic term in the polarization potential than one will have to perform a more thorough stability analysis \cite{tilted-cucumber}.

We can now combine this intuition with the results of \cite{Bena:2014jaa} in order to ascertain whether anti-D3 branes at the bottom of the Klebanov-Strassler (KS) solution \cite{Klebanov:2000hb}\footnote{Which
have a brane-brane-repelling tachyon in the regime of parameters where their backreaction is important \cite{Bena:2014jaa}, and the number of anti-D3 branes is larger than the square of the KS three-form flux.} can still give rise to metastable vacua when polarized into NS5 branes inside the $S^3$ at the bottom of the KS solution \cite{Kachru:2002gs}. As shown in \cite{Bena:2014jaa}, the ``maximum-force'' orientation of these NS5 brane shells is inside the $S^3$, but the ``maximum-tachyon'' orientation is in a different direction. Furthermore, the value of the brane-brane-repelling tachyon inside the ``maximum-force'' plane is exactly zero  \cite{Bena:2014jaa}. It is not hard to see that this NS5 brane will lower its energy by tilting: As one inclines the brane away from the ``maximal-force'' plane, the coefficient of the cubic term decreases like the square of the tilting angle (given by the phase of $z$ in equation (2.11) in \cite{Zamora:2000ha} or equation (3.1) in  \cite{Bena:2014jaa}); however, since this brane does not lay inside the maximum-tachyon plane, the coefficient of the quadratic term will decrease linearly with this tilting angle. Since the two terms are of the same magnitude, the overall potential will decrease as the NS5 brane tilts away from the ``maximum-force'' and settles along another plane that sits between the ``maximum-force'' and the ``maximum-tachyon'' planes. Since the tachyon is exactly zero in the ``maximum-force'' plane \cite{Bena:2014jaa}, and maximal in the ``maximum-tachyon'' plane, it will have a finite nonzero value in this plane. One can then use the result of our analysis to show that this brane will be unstable. Hence, in the regime of parameters where the antibranes backreact our analysis implies that anti-D3 branes polarized into NS5 branes inside the $S^3$ at the bottom of the KS solutions are unstable. 

Thus, even if our investigation does not address in full all the possible decay channels of all possible polarization channels of tachyonic branes, it shows that the most commonly used ones - the anti-D3 branes polarized into NS5 branes at the bottom of the KS solution - are unstable. It is clearly a very important open question to analyze in detail all polarization channels of these antibranes and to find whether there is any metastable one, or whether they are all unstable. It is also important to understand in general whether brane polarization can cure any brane-brane repelling tachyon, or if even the tiniest such tachyon is enough to destabilize all polarized branes. It is also interesting to analyze the polarization of anti-M2 branes in the presence of such tachyons. In \cite{Massai:2014wba} it was found that anti-M2 branes placed in bubbling Lin-Lunin-Maldacena (LLM) geometries \cite{Lin:2004nb} can give rise to metastable vacua in which these anti-M2 branes are polarized into M5 branes. The polarization was induced only by a tachyonic term, and if our analysis extends to that situation it would imply that these metastable anti-M2 LLM vacua are in fact unstable. 

Another interesting direction is to explore in general the fate of the P-wave instability of multiple D3 branes polarized into a {\em single} D5 brane inside  a tachyonic $AdS_5 \times S^5$ solution sourced by these D3 branes. Our calculation gives the polarization potential of a probe in a general tachyonic background and, as explained in \cite{Polchinski:2000uf}, this probe potential can be used to find the potential describing the polarization of all the D3 branes sourcing the solution. When the D3 branes polarize into multiple five-brane shells, the P-wave tachyonic mode we find above corresponds to shifting the centers of these five-brane shells away from each other. However, if one assumes that our calculation of the polarization potential extends to the configuration in which all the D3 branes are polarized into a single five-brane, it is unclear what this P-wave tachyonic mode corresponds to.\footnote{This discussion does not apply to backreacted antibranes, for which we have only shown that a brane-brane-repelling tachyon is present in a regime of parameters in which the dipole charge of the polarization shell is greater than one.}

Our analysis explores the effects of the brane-brane-repelling tachyon on the stability of brane polarization shells using the Born-Infeld action of these shells. However, brane polarization can also be described using the degrees of freedom of the branes that polarize \cite{Myers:1999ps, Polchinski:2000uf}, and this description is valid in the regime of parameters where the backreaction of these branes is not important\footnote{Although the existence of a brane-brane-repelling tachyon was only established when $g_s N \gg 1$ \cite{Bena:2014jaa}, the technology now exists \cite{Michel:2014lva} for studying whether non-backreacting antibranes also have tachyons; a simple exploration of the $SO(6)$ transformations of the various terms that can appear in the brane-effective-action \cite{Michel:2014lva} indicates that a brane-brane-repelling tachyon might as well be present in this regime.} ($g_s N \ll 1$). In the vacua where the branes are polarized, the $N \times N$ scalar fields living on the branes become non-commutative. It is not hard to see that upon deforming the Lagrangian of ${\mathcal N}=4$ SYM with a tachyonic traceless boson bilinear term of the form $m^2 {\rm Tr}(-\Phi_1^2 - \Phi_2^2 - \Phi_3^2 +\Phi_4^2 + \Phi_5^2 + \Phi_6^2)$ one can easily find vacua where these fields become non-commutative 
\begin{equation}
[\Phi_1, \Phi_2] \sim m ~\Phi_3 \quad + \quad \textrm{cyclic permutations,} 
\end{equation}
and these ``Higgs'' vacua describe the polarization of the D3 branes into D5 branes. Hence, from the point of view of the theory on the branes, a tachyon that causes a runaway behavior on the Coulomb branch (when the scalars commute) can still give rise to Higgs-branch vacua. The question is then whether these Higgs-branch vacua are metastable, or are tachyonic, and also what is the matrix equivalent of the P- and D-wave instabilities we have found. More generally, to prove that the Higgs-branch vacua are metastable one would have to show that the ${\cal O}(N^2)$ possible perturbations of this vacuum give rise to an ${\cal O}(N^2 \times N^2)$ mass matrix that has only positive eigenvalues. At first glance this appears rather unlikely, but if such a metastable vacuum exists it would have quite extraordinary implications for our understanding of vacua in tachyonic non-Abelian gauge theories.

\acknowledgments{We would like to thank Johan Bl{\aa}b{\"a}ck, Ben Freivogel, Mariana Gra\~na, Praxitelis Ntokos, Silviu Pufu and David Turton for interesting discussions.
This work was supported in part by the ERC Starting Grant 240210
{\em String-QCD-BH}, by National Science Foundation Grant
No.~PHYS-1066293 (via the hospitality of the Aspen Center for Physics)
by the John Templeton Foundation Grant 48222 and by a grant from the
Foundational Questions Institute (FQXi) Fund, a donor advised fund of
the Silicon Valley Community Foundation on the basis of proposal
FQXi-RFP3-1321 (this grant was administered by Theiss Research).}

\appendix

\section{Explicit expressions for the coefficients of the tachyonic potentials.}

\label{sex:Appendix}

We give below the full expression of the coefficients $v_0 (\omega)$, $v_2 (\omega)$ and $b_{1,2,3} (\omega)$ that appear in the P-wave and D-wave brane polarization potentials (\ref{eq:V-onshell}) and (\ref{eq:elipsoid-pot}):

\begin{eqnarray}
v_0 (\omega) &=& - \frac{1}{128} \frac{ \left(3 + 8 \omega^2 \right) \left( 81 + 360 \omega^2  + 128 \omega^4 \right) + \left(81+432 \omega^2+448 \omega^4 \right) \sqrt{9+32 \omega^2}}{\left( 3+\sqrt{9+32 \omega^2} \right)^2}
 \\
v_2 (\omega) &=& - \frac{4}{3} \omega^2 \frac{ \left( 3 \left( 27 + 88 \omega^2 \right) + \left( 27 + 40 \omega^2 \right) \sqrt{9+32 \omega^2}\right)}{\left( 3+\sqrt{9+32 \omega^2} \right)^3}  \label{eq:v_2app}
\end{eqnarray}
\begin{eqnarray}
b_1 (\omega) &=&  \frac{1}{45} \frac{ \left(  \left( 567 + 2160 \omega^2 + 896 \omega^4 \right) + \left( 189 + 384 \omega^2 \right) \sqrt{9+32 \omega^2}\right)}{\left( 3+\sqrt{9+32 \omega^2} \right)^2} 
 \\
b_2 (\omega) &=&  \frac{1}{180} \frac{ \left(  \left( 1053 + 3240 \omega^2 - 256 \omega^4 \right) + \left( 351 + 456 \omega^2 \right) \sqrt{9+32 \omega^2}\right)}{\left( 3+\sqrt{9+32 \omega^2} \right)^2} 
 \\
b_3 (\omega) &=&  \frac{1}{180} \frac{ \left(  \left( 324 + 4320 \omega^2 + 8192 \omega^4 \right) + \left( 108 + 1248 \omega^2 \right) \sqrt{9+32 \omega^2}\right)}{\left( 3+\sqrt{9+32 \omega^2} \right)^2} 
\end{eqnarray}

\bibliographystyle{utphys}
\bibliography{Draft-cucumber}

\end{document}